# A biophysical model of prokaryotic diversity in geothermal hot springs

Anna Klales, James Duncan, Elizabeth Janus Nett, and Suzanne Amador Kane

Physics Department, Haverford College, Haverford PA 19041

Recent field investigations of photosynthetic bacteria living in geothermal hot spring environments have revealed surprisingly complex ecosystems, with an unexpected level of genetic diversity. One case of particular interest involves the distribution along hot spring thermal gradients of genetically distinct bacterial strains that differ in their preferred temperatures for reproduction and photosynthesis. In such systems, a single variable, temperature, defines the relevant environmental variation. In spite of this, each region along the thermal gradient exhibits multiple strains of photosynthetic bacteria adapted to several distinct thermal optima, rather than the expected single thermal strain adapted to the local environmental temperature. Here we analyze microbiology data from several ecological studies to show that the thermal distribution field data exhibit several universal features independent of location and specific bacterial strain. These include the distribution of optimal temperatures of different thermal strains and the functional dependence of the net population density on temperature. Further, we present a simple population dynamics model of these systems that is highly constrained by biophysical data and by physical features of the environment. This model can explain in detail the observed diversity of different strains of the photosynthetic bacteria. It also reproduces the observed thermal population distributions, as well as certain features of population dynamics observed in laboratory studies of the same organisms.

87.23.-n

## I. INTRODUCTION

 Life on earth is surprisingly resilient, exhibiting a wealth of organisms capable of inhabiting regions of extreme heat, cold and pressure, as well as harsh chemical environments with extremes of salinity, acidity or alkalinity.  Of these, geothermal hot springs environments have attracted considerable attention because they represent an ideal opportunity to study microbial evolution in action, providing a model for the evolution of early life on earth and possibly other planets. [1-3] [4]

In this paper we consider studies of the ecology of communities of photosynthetic bacteria living along thermal gradients in alkaline hot springs, a system for which extensive genetic and biophysical data are available.  These systems offer an opportunity to model the interplay of population biology and evolution with the physical environment:  a limited and well-defined number of species interact within a simple, well-characterized environment that appears to determine fully the evolutionary and population dynamics.  To explain the observed diversity and distribution of distinct strains of bacteria in space and temperature, we performed computational studies using a combination of population biology, artificial life and biophysical models and data; we also reanalyzed the field microbiology literature to point out certain universal features of these systems.

## II. BACKGROUND

Terrestrial geothermal hot springs have been present since the early evolution of the earth, and they are found throughout the world in the modern era. Their surface water temperatures range from boiling or superheated at their sources to ambient temperature at their shores, while their chemical composition can feature high acidity or alkalinity and dissolved minerals, methane, and hydrogen sulfide, which can serve as nutrients for their associated biota. [5] All three domains of life, Archaea, Bacteria and Eukarya, are represented in these systems, although most thermophiles (organisms adapted to temperatures of 50 to 90$^o$C) are predominantly bacteria, and most hyperthermophilic organisms (those with optimal temperatures for growth > 90$^o$C) are archaea. [6] Such organisms have evolved a number of biochemical mechanisms to avoid thermal damage, only some of which are presently understood. [7]

Even in the fluid flow rates found in hot springs, communities of prokaryotic thermophiles can achieve a stable geometry by forming microbial mats, structures in which individual cells are joined into a community by an extracellular polymeric matrix. [8, 9] Microbial mats allow prokaryotes to achieve a stable spatial location both horizontally and vertically, allowing them to position themselves at defined regions of temperature, light intensity, salinity and nutrient concentration. Consequently, hot springs often exhibit distinctively colored bands, consisting of different microbial species living at preferred zones of average temperature. (Fig. 1)

Understanding these prokaryotic communities' structure can provide insights into an important epoch in early life. Considerable evidence indicates that microbial ecosystems similar to these have been present for approximately 3.5 Gyr, since early in the evolution of life. [10] Their structure can be studied through features in the fossil record, including stromatolites (organosedimentary geological structures produced by microbial mats, especially cyanobacteria), microfossils (including fossilized remains of single cells) [11] and biomarkers (chemical tracers) preserved in sedimentary rocks. [12] For example, biomarker methods such as isotope analysis and the presence of characteristic lipids can be used to deduce the presence of particular species of bacteria, archaea and eukaryotes in geological deposits from periods corresponding to the early evolution of life.

In this study, we are concerned in particular with microbial mats of photosynthetic bacteria, which are observed over a temperature range of approximately $45^{o}C$ to $73^{o}C$. The upper temperature limit is determined by the lethal temperature for the highest temperature strains for each species. At temperatures below $45^{o}C$, the observed microbial mats are increasingly depleted by grazing due to predation, which can be ignored at higher temperatures, as well as competition from non-thermophilic species. [13] Study of these systems is greatly simplified by the limited microbial diversity observed over this temperature range. The absence of predators also makes this zone an attractive analogue for life on early Earth, when multicellular organisms were absent. Microbial mats in this temperature range often consist of vertically-organized communities in which the photosynthetic cyanobacterium *Synechococcus lividus* Copeland (hereafter *Synechococcus*) can live as a top layer while the photosynthetic green nonsulfur-like bacterium *Chloroflexus aurantiacus* lives as an undermat. This geometry results because *Chloroflexus* utilizes bacteriochlorophylls *a* and *c*, (BChl *a* and *c*) chromatophores with a different absorption spectrum from the chlorophyll *a* (Chl *a*) characteristic of cyanobacteria, so it can make use of sunlight at wavelengths poorly absorbed by the mat's top layers.  Thus, we can assume that these two species do not significantly compete for resources within the topmost layer.  Given their complementary utilization of sunlight and ecological niches, there is evidence that significant symbiotic interactions may occur. [5, 14]

In addition to the simple species composition of these systems, such communities consist of ecosystems whose detailed population makeup is defined by a single variable: local temperature along a thermal gradient.  When first discovered, the photosynthetic bacteria in these systems were expected to exhibit strains with optimal temperatures for growth and photosynthesis matched to the local temperature of each region along the thermal gradient, because of the short half-time for a bacteria species to outcompete a less fit competitor in the field.[15]  Instead, multiple strains of bacteria are found at each point along the gradient, with a variety of optimal temperatures that encompass values well above or below the local environmental temperature.  Indeed, it is common to find multiple temperature strains with overlapping territories existing in equilibrium for long periods of time.  Various field studies have mapped out both the detailed population distribution of these organisms and their temperature response to gradients.  Here we seek a more detailed explanation for this surprisingly diverse and stable community structure.

Thermal gradients that support microbial mat communities occur in various settings:  between the center of geothermal hot springs and their edges, along effluent channels that drain hot springs, as well as along the vertical near the ocean's surface.[5]  In the case of effluent channels and the edges of hot springs, the temperature gradient can be modeled as having a linear profile over most of its range [5, 16, 17] and this also holds in other scenarios of interest, such as shallow-water hydrothermal vents [18] and water columns in the oceans. [19] The distances over which these thermal gradients exist are large compared to individual organisms, typically many meters in extent. In hot springs, this temperature variation is not accompanied by variations in sunlight intensity or spectral content; the composition of unstable chemical constituents such as hydrogen sulfide has been observed to closely track the thermal gradient. [17] While some species of photosynthetic bacteria are highly mobile and can position themselves to find their optimal environmental conditions, the species considered herein have not been found to migrate horizontally so as to track temperature variations. [20, 21] Seasonal temperature variations might be expected to be a major factor in defining the thermal environment, as well as a logical selective pressure that favors the retention of different temperature strains; hot springs also can exhibit short-term temperature variations due to water surges over timescales as short as a few minutes. However, little variation in average temperature ($\leq 3^{\circ}$C) was observed seasonally in the cases considered. While some systems exhibited considerable periodic, shorter-term temperature fluctuations, these were found to have communities similar to those lacking significant thermal variation. [22] [23] [24]

Because of the difficulty in delineating distinct species for organisms, such as bacteria, that reproduce asexually, we use the term *strain* (or equivalently *ecotype*) to distinguish distinct variants within a species. [2] Recently, the study of bacterial temperature strains has been greatly facilitated by analyses of the sequence of the small subunit ribosomal RNA gene, a technique particularly valuable for thermophilic species that have proven difficult to study by cell culture methods. [25] [26] In particular, such methods have allowed the mapping of unexpectedly diverse lineages of thermophiles within hot springs environments.[14, 24, 27-30] (In other work, high levels (97%) of similarity in 16S ribosomal RNA (rRNA) sequences can correspond to strains with very different environmental preferences, illustrating that such sequence similarity sets only a lower bound on diversity, as it does not rule out subpopulations with similar rRNA but different physiological properties.[19]) Phylogenetic studies also have shown that community structure and species diversity in the hot spring systems considered here are determined primarily by water temperature, rather than chemical composition, pH or other factors.[13, 24, 31] In different settings, similar population distributions can result from salinity gradients[32] or variations in the intensity of sunlight.[19] In addition to such phylogenetic studies, strains of bacteria identified by such techniques can be cultured and studied to determine such biophysical parameters as the temperature dependence of their birth rates, photosynthetic rates and other metabolic parameters. The recent sequencing of entire *Synechococcus* genomes also promises to further elucidate the genetics of these organisms.[27]

Several groups have developed evolutionary ecology explanations for the present diversity of cyanobacteria and other prokaryotes in these settings. Possible explanations include the adaptive radiation of higher temperature strains from a progenitor lower temperature strain, and the idea that each spring contains only that subset of all possible species in the larger environment that can survive the local conditions, such as elevated temperature.[33] For example, Miller and Castenholz have used results of phylogenetic analyses from Hunter's Hot Springs (Oregon) cyanobacteria to show that more thermophilic strains likely evolved from a single, more temperate ancestor.[29] However, to our knowledge, no work has been done yet to explain the spatial population distribution of temperature strains in these physical environments or the observed distribution of strains with different optimal temperatures. Any theory to explain these observations needs to expand beyond the traditional boundaries of population dynamics models (that usually include only time but not spatial dependence) to include their interplay with the crucial biophysical characteristics of the bacteria and the physical environment that distinguish these systems.

Our model of these systems incorporates elements of the Daisyworld model originally proposed by Watson and Lovelock [34] to illustrate the potential for biological homeostasis as a means of regulating planetary temperature. More recent studies of Daisyworld models have examined natural selection, [35] pattern formation, [36] and desert formation. [37] In our study, Daisyworld provides a framework for combining population biology models [38] with biophysical data about the optimal temperature dependence of bacterial birth rates and enzymatic activity and information about the physical environment (primarily temperature, chemical composition and light intensity). Using this approach, we probed possible explanations for the observed data. For example, seasonal variations or short-term temperature variations might favor a wider range of temperature adaptations because genetic diversity within the population might increase the fitness of an entire species in the face of environmental fluctuations. However, the observed level of diversity might result instead because competition between different thermal strains favors an optimal distribution of strains with distinct optimal temperatures.

# III. MATERIALS AND METHODS

## A. Computer simulations and data analysis

Our computations were performed using programs written in C++ running under the Visual C++ (Microsoft Corporation, Seattle WA) and Dev C++ (Bloodshed Software, Open Software Foundation, www.bloodshed.net) programming environments.  All software was run on Dell Optiplex 745 personal computers with Intel Dual Core 2.66 GHz processors and 4 GB memory running Windows XP.

Data from previously published research was digitized using SigmaScan Pro 5.0 image analysis software (SPSS, Inc., Chicago IL).  All curve-fitting and other data analysis was performed using Origin version 7.03 (OriginLab Corp., Northhampton MA).

## B.  Model equations and parameters

The growth of the bacterial strain populations in our model is governed by difference equations describing the spatiotemporal dynamics of population density on a one-dimensional spatial grid.  Each location along the 1D grid represents a point along a linear temperature gradient, $T(z)$, from 25 to 80$^o$C extending over N = 500 spatial grid points in the $z$-direction.  For example, for a gradient 5 meters long, the approximate size ($\Delta z$) of each cell would correspond in the physical system to approximately 1 cm, so each cell should be thought of as encompassing a large population of bacteria in general.  The light intensity was assumed to be uniform in space and time except for studies of seasonal variation, as explained below.[39]  Our temperature is in units of ºC, but the light intensity is in arbitrary units.

We define $\rho_i(z,t_n)$ as the bacterial population density of the $i$th temperature strain at location $z$ and $n$th time step $t_n$, and $\Delta\rho_i(z,t_{n+1})$ as the change in density between the $n$th and $(n+1)$th time step.  Different temperature strains are indicated by the index $i$, which ranged from 1 to $N_s$, the total number of strains with different temperature optima.  We also define the total population density summed over all temperature strains as

$\rho_t\left(z,t_n\right)=\sum_{i=1}^{N_s}\rho_i\left(z,t_n\right)$.  Given the scale of the spatial grid, these population densities were not expected to agree numerically with those in natural settings, but are meant rather to simulate the proportional distribution across different temperatures.  Each bacterial population follows a difference equation derived from the usual Lotka-Volterra model for population dynamics [38]:

$$\Delta\rho_i\left(z,t_{n+1}\right)=\rho_i\left(z,t_n\right)\left(\beta_i\left(z\right)\left(1-\frac{\rho_i\left(z,t_n\right)}{K_i\left(z\right)}\right)-\lambda_i\right)-\mu\cdot\rho_i\left(z,t_n\right)\sum_{j\neq i}\rho_j\left(z,t_n\right) \qquad (1)$$

where $\beta_i$ is the temperature-dependent  birth rate and $\lambda_i$ is deathrate.  The second term serves to downregulate population growth of the $i$th strain when the population exceeds the temperature-dependent carrying capacity, $K_i$) the maximum value supported by environmental conditions.)  Since bacterial populations reproduce asexually, data for measured birth rates are reported in terms of temperature-dependent doublings per day, $1/\tau_2$, or equivalently birth rate $\beta=\Delta t\ln 2/\tau_2$, where the time step, $\Delta t=t_n-t_{n-1}=1$ hr typically for these simulations.  The model showed little sensitivity to variation of $\Delta t$ over an order of magnitude.  The final term in (1) is meant to model the effects of competition for resources between strains using a competition factor, $\mu$, assumed for simplicity to be the same for all cyanobacteria strains. For the sake of our model, we assumed no competition between strains that had distinct mRNA sequences, yet identical temperature preferences since interactions between strains with overlapping temperature strains were already incorporated into the carrying capacity term. Only strains with different temperature optima have non-zero competition terms, and those populations were merged if their values of $T_o$ agreed within error bars.

We modeled how the thermal gradient would affect population dynamics by incorporating biophysical characterizations of enzymatic activity and metabolism. In addition to introducing physical and spatial/thermal dependence into our model, this also allowed us to use known values for all of the parameters in (1), except for the competition factor, which was obtained by comparison with actual population distribution data. The carrying capacities and birth rates in prokaryotes always exhibit an optimal temperature, $T_o$, at which the enzymatic activities and birth rates are at a maximum. [6] This is due to two effects: at high temperatures, thermal denaturation destabilizes the structure of key enzymatic binding sites, while at suboptimal temperatures, the Arrhenius law gives an exponential increase of enzymatic activity with temperature. These competing behaviors result in the curves shown in Fig. 2 (a)-(d), which show the characteristic temperature dependence of birth rate (Fig. 2(a)-(b)) and photosynthetic rate (Fig. 2(d)) for photosynthetic bacteria in hot springs. For each strain, there is a lowest temperature, $T_L$, above which the birth rate becomes nonzero, as well as a lethal temperature, $T_H$, above which the bacteria cannot live or reproduce if subjected to this condition for several hours. [40]

In our model, we simplify this behavior and represent both temperature preferences using the functional forms for the birth rates derived from the Daisyworld research of Lovelock and Watson. [34]  Daisyworld models the birth rate of a "daisy" (model cyanobacterium-like organism) as a function of temperature with an inverted, parabolic shape, with different curvatures on the high and low temperature side of the optimal temperature, $T_o$.  (Fig. 2(a))  (The exact shape of this distribution is not important to our main conclusions, as was established by using different functional forms for the high and low temperature roll-offs.)  Using a parabolic functional form, the birth rate temperature dependence was approximated by:

$$\beta_i\left(z\right) = \beta_o\left(1.0 - \upsilon_{i\beta}\left(T\left(z\right) - T_{oi}\right)^2\right) \qquad (2)$$

where $T_{oi}$ is the optimal temperature for growth for the $i$th temperature strain, $\beta_o$ is the maximum birth rate at $T_{oi}$ and the normalization constant $\upsilon_{i\beta}$ for each organism is determined by the maximum and minimum temperatures for growth, $T_{H\,i}$ and $T_{L\,i}$, such that

$$\upsilon_{i\beta} = \frac{1}{\left(T_{ji} - T_{io}\right)^2}, \qquad (3)$$

where $T_{ji}$ is either $T_{Hi}$ (for $T\left(z\right) > T_{oi}$) or $T_{Li}$ ($T\left(z\right) > T_{oi}$) for the $i$th temperature strain. Similarly, each organism's carrying capacity has the same functional form to capture the temperature dependence of the rate of photosynthesis:

$$K_i\left(z\right) = K_o \times L\left(z, t_n\right) \times \left(1.0 - \upsilon_{iK}\left(T\left(z\right) - T_{oi}\right)^2\right) \qquad (4)$$

where the values of $T_{oi}$, $T_{Hi}$ and $T_{Li}$ were used to compute $K_i$, and its normalization factor $\upsilon_{iK}$ for each strain can in general differ from those for the birth rate.  The maximum carrying capacity $K_o$ was set to a uniform value (here 100 in arbitrary units for population density per spatial grid point), to reflect the fact that observations do not show large variations in the maximum photosynthetic rate between temperature strains. The effect of varying light levels is captured in the function $L(z,t_n)$.

All parameters for the temperature-dependent birth rates and carrying capacities were derived from experimental data, as explained below.[a] The total rate $\lambda_i$ has contributions from the death of individuals, as well as losses due to organisms being washed out from the mat. Microbial mats often persist for long times without deterioration, so a death rate of $\lambda = 0$ was used in our simulations.[1] For *Synechococcus* microbial mats denied all light, an upper-bound on the death rate was measured to be $\lambda = \Delta t$ ln 2 /$\tau_{1/2}$ = (2 hr) ln 2/(40 hr) = 0.035 hr$^{-1}$; our simulations were run using both values, with negligible effect on final populations.[41]

Where appropriate to model either seasonal or shorter-term fluctuations, temperature and light variations were modeled as sinusoidal fluctuations of sunlight intensity $L(z,t_n)$ and the ambient temperature $T_f(z, t_n)$ with a period of $\tau_f$. The highest temperature limit was fixed by the presumably constant hot spring source. For seasonal variations $\tau_f$ was set to 1 year, while shorter-term variations were given a period of 5 minutes, in agreement with empirical data for water surges in some hot springs.[42] Values on the order of a few degrees to 10 $^{\circ}$C to 15 $^{\circ}$C peak to peak were observed empirically in shorter-term variations.  The measured variation in ambient temperature at the relevant latitude for our studies over a year is 19.8 $^{o}$C peak to peak on average. [39]  While t the amplitude of temperature fluctuations was varied between 0 and 50ºC in the simulations, values greater than 10 $^{o}$C represent catastrophic scenarios in which temperature surges might be expected to eliminate some temperature strains on occasion.

---

[a] The supplementary materials contain a sample C++ program used to predict population densities, including the effect of predation and characteristic values of all parameters.

Given data indicating that these cyanobacteria do not move laterally to track temperature, we did not consider the effect of bacteria migrating into adjacent cells, so the behavior at each spatial location is independent of those at other locations. [21]  We also simulated the evolution in time for these systems under various scenarios of thermal fluctuations.  Growth curves for each strain and for the total population initially exhibited followed the approximately exponential growth as expected, with a gradual leveling off of the total population density as the carrying capacity is reached, in agreement with experimentally determined bacterial growth curves for these systems. [43]  For the experimentally determined parameters used, the total population density typically reached 98% of equilibrium values after approximately 100 time steps.

## C. Modeling the effect of predation

To include the effect of predators, our model needed to be modified only at the lower end of the temperature range, consistent with studies by Wickstrom and Castenholz of the effects of grazing by the predominant predator species. [13]  These studies determined the effect of thermophilic ostracods, microscopic crustaceans that feed on the cyanobacterial mats at temperatures approaching the ostracods' lethal temperature by swimming up the thermal gradient while grazing, then drifting safely downstream once they go into a temperature-induced coma. By this means, the ostracods virtually eliminate the cyanobacterial population below approximately 45°C and reduce its value up to approximately 50 °C. Quantitative measurements of ostracod population density, $\rho_o$, vs. temperature were available from these studies; this data was digitized from Fig. 7 from Ref. . [13] (Fig. 3) In our simulations, the ostracod population data were modeled using a linear fit for temperatures above 47.5°C and a quartic-polynomial for lower temperatures. Fig. 12 from Ref. [13] was also analyzed to determine that a linear relationship held between the measured ostracod density and the loss of cyanobacterial biomass, justifying our use of a competition term like that in (1) to model ostracod predation. As a result, we modified (1) to include a new term:

$$\Delta \rho_i\left(z, t_n + 1\right) = \rho_i\left(z, t_n\right)\left(\beta_i(z)\left(1 - \frac{\rho_i\left(z, t_n\right)}{K_i(z)}\right) - \lambda_i\right) - \mu \, \rho_i\left(z, t_n\right)\sum_{j \neq i}\rho_j\left(z, t_n\right) - \mu_p \rho_i\left(z, t_n\right)\rho_o\left(z, t_n\right). \quad (7)$$

Since the new predation competition factor, $\mu_p$, could not be determined independently from empirical data, its value was chosen to by comparing simulated with measured *Synechococcus* population density distributions.

## IV. RESULTS

### A. Re-analysis of the experimental data

#### 1. *Measured distribution of optimal temperatures along the thermal gradient*

We first re-analyzed data from multiple studies of photosynthetic bacteria in hot springs settings throughout the world. Bringing together these results elucidated several shared features of the data, such as similarities in the temperature distributions in different locations, and allowed us to extract parameters needed for the simulations using (3) and (4). In each case studied, many different strains were found, but only a limited number of different temperature optima, $T_o$. *Synechococcus* populations from different geographic locations with similar thermal behavior were found to be genetically distinct. For example, none of the Oregon strains identified in Ref. [29] were identical to those found in Yellowstone National Park in Ref. [23]. Since we were interested in thermal behavior, we only focused on strains with distinct temperature dependence here.

Allewalt *et al.* [44] were able to obtain the temperature dependence of birth rates and photosynthetic rates for *Synechococcus* strains found along a thermal gradient in Octopus Springs, Yellowstone National Park. They were able to measure temperature dependences of birth rates for their genetically distinct strains A, B, B', B'' and B'''; they were unable to cultivate their strain A', but noted it has similar characteristics to Group IV strains in Ref. [29]. In the one case (strain A') where data did not exist for the temperature dependence of the photosynthetic rate, we were able to use the same range, $T_H$-$T_L$, as found for the other, well-characterized strains, and estimate $T_o$ for photosynthesis by noting there was little variation between measured strains in this case.[44]

Bauld and Brock [45] performed studies of both *Chloroflexis* and *Synechococcus* along an effluent channel from a pool in Lower Geyser Basin (Mushroom Springs), Yellowstone National Park, in summer. The temperature dependence of the photosynthetic rate was measured for four strains of *Chloroflexis* sampled at 45, 50, 60 and 72 °C (Fig. 2(d)), and used to obtain values for $T_o$.

Peary and Castenholz [46] measured the birth rate dependence on temperature for multiple *Synechococcus* strains from Hunter's Hot Springs in Oregon sampled at a variety of temperatures (45, 48, 53, 55, 60, 66, 71 and 75 $^{\circ}$C) (Fig. 2 (c)). This setting did not experience significant periodic temperature fluctuations. It is notable that the actual sampling temperatures varied and were not uniformly spaced; also, comparison of $T_s$, the local environmental temperature at which samples were obtained, and $T_o$ revealed that $T_s \neq T_o$ for the two cases in the literature where the sampling temperature were reported, ruling out the obvious interpretation that optimal temperatures were simply a reflection of the sampling methods.[45, 46] A later rRNA sequence analysis of cyanobacteria from this setting by Miller and Castenholz found a similar distribution of temperature strains. [29]

Given these field data results, we were able to use measured dependencies for birth rate and photosynthetic rate upon temperature wherever possible in our simulations for the parameters in (3) and (4). To analyze the distribution of temperature strains from this data, we first determined each strain's $T_o$, as explained below. In order to understand how these values were distributed, we next computed $\Delta T$, defined as the difference in optimal temperature between temperature strains with the closest distinct values of $T_o$. (See Fig. 2(a).) Strains for which the optimal temperatures were indistinguishable within error bars were grouped together for the sake of these calculations. In each of the cases studied, we were able to obtain distinct values of $T_o$ for four distinct temperature strains: one set for *Chloroflexus* [45], and the three sets for *Synechococcus* from Refs. [29 44 46]. (Fig. 2(b)-(d))

We measured $\Delta T$ from data for multiple strains by first measuring the optimal temperatures using two techniques that agreed within experimental uncertainties: 1) the maximum birth rate (or photosynthetic rate) was taken directly from the peak value measured; and 2) a parabolic fit was used to find the maximum. Next, the difference between strains with the closest optimal temperatures was computed. In yet another check, we determined $\Delta T$ as the difference between the middles of each temperature range (i.e., the average of $T_H$ and $T_L$) for successive temperature strains. We also determined the entire temperature range for each dataset (defined as $T_H - T_L$), to see whether controlling for the width of the birth rate vs. temperature curve would affect our data; in every case studied, the average range agreed within uncertainties for each organism considered.

## B. *Measured cyanobacterial population distributions along the thermal gradient*

We next compared estimated distributions of cyanobacteria populations from four different studies in four different geographic locations, and for one location for *Chloroflexus*. Similar results have been seen for other organisms also distributed along thermal gradients, including *Thermus aquaticus* [17 47] and Icelandic blue-green algae[48], albeit with different thermal ranges in the first and last examples. The population density data as a function of collection temperature were drawn from a variety of techniques, including UV-visible spectra of photosynthetic pigments (chlorophyll-*a* (for *Synechococcus*)) and bacteriochlorophyll *a* and *c* (for *Chloroflexus*)) and in situ cell counts. Where noted, we include the time of year of collection, as well as any descriptions of the variations in the temperature at the collection sites.

T.D. Brock [22] reported on cell culture studies of *Synechococcus* in effluent channels from hot springs in Lower Geyser Basin (Mushroom Springs), Yellowstone National Park. This work examined bacteria from regions of linear temperature gradients that were uniform across the effluent channel, and that exhibited laminar water flow and uniform microbial mat formation. The total temperature variation at fixed locations along the thermal gradient were measured throughout the year and found to vary by $\pm$ 3 to 5 $^o$C. Quantitative spectroscopic measurements of Chl $a$ absorbance were used to measure the distributions of *Synechococcus* vs. temperature in the summer. (Fig. 4)

Bauld and Brock [45] determined the population distributions of photosynthetic bacteria along a thermal gradient in Lower Geyser Basin (Mushroom Springs), Yellowstone National Park. Temperatures fluctuated by $\pm2.5^o$C (full range) during the period studied. Population abundances (relative concentrations per unit area) were determined for each species (*Chloroflexis* and *Synechococcus*) using spectrophotometric measurements of BChl $a$ and $c$ and Chl $a$ absorbance, respectively. (Fig. 4)

Sompong *et al.* [24] measured cyanobacterial distributions for a variety of hot springs in Thailand, at temperatures within the ranges 30-40, 40-50, 50-60, 60-70 and 70-80$^o$C, for February and August. The abundance of *Synechococcus* as a function of temperature, summed over all strains, was measured by cell counts using an optical microscope. (Fig. 4)

Only Ferris and Ward [23] measured data reflecting population abundances of *Synechococcus* for both total population, $\rho_t$, and individual bacterial strains, $\rho_i$. In this work, denaturing gradient gel electrophoresis (DGGE) data was used to determine the presence of various strains of cyanobacteria, as well as green nonsulfur bacteria-like and sulphur bacteria-like organisms, at 5 different temperatures along a thermal gradient in Octopus Springs, Yellowstone National Park, at four distinct seasons. No mats with these organisms were found below 42 and above 74 $^{\circ}$C. [14] Local temperatures were observed to fluctuate by 1 to 12$^{\circ}$C over a 2-minute period due to water surges. The average temperature at two fixed locations were reported to be constant from season to season, however.

To obtain approximate population abundances for each temperature strain, we digitized the DGGE data from Ref. [23] to obtain the grey-scale intensity within each band corresponding to a distinct temperature strain, then integrated each band's total intensity. Measurements of the grayscale intensity levels of the gel's background were used to normalize the data from different gels to compensate for scaling due to different exposures. Since DGGE profiles were not linear with concentration in the regime studied, due to PCR amplification factors, these results only show overall trends, and should not be interpreted as exact gene (and hence strain) abundances. From dilution data provided by the authors, we infer that these profiles should show less variation for higher abundances than would occur in the actual populations, so we would expect this data to correspond to a flattened distribution near the optimal temperature relative to the actual abundance, as is indeed observed relative to other methods. (Fig. 4) (However, other studies have validated comparisons of DGGE estimates of population density using accompanying direct cell counts. [32]) There were seasonal variations in DGGE intensity for each strain, but we present only the data for spring (March) here; similar results were obtained for total cyanobacteria population distributions for summer (June).

**B. Modeling genetic diversity**

*1. Measured and computed distribution of optimal temperatures*

We now present an analysis of the studies that have provided comprehensive biophysical data on the distribution of optimal temperatures along a thermal gradient.[29, 44-46] (Fig. 2 (b)-(d)) Two features of the measured distributions of optimal temperatures stand out. First, for the fixed range of temperatures along which these organisms can live on the gradient, four distinct temperature strains exist in each case. Second, the optimal temperatures are spaced relatively uniformly with similar values of $\varDelta T$ for all cases studied. For the nine distinct values of temperature spacing found from the measured data, the average value was $\varDelta T = 5.2 \pm 0.7\ ^{\circ}\text{C}$.

These observations can be explained by a simple argument to predict the expected optimal number of strains and $\varDelta T$, using only a few assumptions borne out by the empirical data. We noted that the upper temperature bound for growth for the strain with the highest optimal temperature was 73 $^{\circ}\text{C}$, the lethal temperature for all the systems studied. Also, at 28 $^{\circ}\text{C}$ on average all birth rates reported either were measured to be zero or were extrapolated to go to zero. Combining these facts with the observation that $\varDelta T$ was uniform, we can see from Fig. 2(a) that the following relation must hold:

$$73^{\circ}\text{C} - 28^{\circ}\text{C} = \left( N_s - 1 \right)\varDelta T + \left( T_{o1} - T_{L1} \right) + \left( T_{HN_s} - T_{oN_s} \right) \tag{8}$$

The data in Fig. 2 were used to compute that, on average, $\left( T_{o1} - T_{L1} \right) = 18.1^{\circ}\text{C}$ and $\left( T_{HN_s} - T_{oN_s} \right) = 8.7^{\circ}\text{C}$. Setting $N_s = 4$, and using these values, we could obtain a prediction that $\Delta\text{T} = 6.1 \pm 1.8\ ^{\circ}\text{C}$, which agrees within error bars with the observed spacing.

We next used our computational model to obtain a value for the competition parameter, $\mu$, for different thermal strains. The rationale was that there would be an optimal number of strains along the thermal gradient for each $\mu$, and this should agree with the observed value of $N_s = 4$ for the actual value of $\mu$. We computed the predicted total population summed over all strains, $\rho_t$, at equilibrium as a function of different values of $N_s$ and $\mu$, using (1) with birthrate parameters determined from Ref. [44], an average peak birth rate at $T_o$ of $\beta_o = 0.58$ (the average value for all strains) and the average values for $T_H$ and $T_L$ mentioned earlier. Fig. 6 illustrates how for this range of temperatures, our model indeed predicts that there exist values of $\mu$ for which different values of $N_s$ are optimal. Plots of our model data revealed that the observed values of $N_s = 4$ corresponded to $\mu = 0.0046 \pm 0.0008$. Values of $\mu$ within this range were subsequently used in predicting the total bacterial population vs. temperature, and its breakdown by temperature strain.

### 2. Population distribution along the thermal gradient

We next considered as a whole the literature on measured photosynthetic bacteria population density distributions along hot spring thermal gradients. Figs. 4 (a) and (b) compares results from different studies of the temperature dependence of the total population density taken in different years at different geographic locations, with the peak values normalized to one for each study. This comparison reveals that there is a very high level of agreement in population distributions for these different studies, comprising populations from four studies of genetically-distinct populations of *Synechococcus* (Fig. 4), indicating that specific genetic strain, geographic location and exact spatial dimensions of the thermal gradient are not important. Indeed, within error bars of temperature and population density, all data sets for net population density exhibit the same temperature distribution between the lethal temperature for cyanobacteria (approximately 73$^{\circ}$C) and approximately 42$^{\circ}$C, where predators and other lower-temperature organisms prey on the cyanobacteria population. (One set, obtained by Brock [22], shows a larger population density at low temperatures, presumably due to either to reduced grazing pressure or inclusion of other photosynthetic bacterial species in this location.) It is also apparent that measurements using different techniques yield very similar data for the same organisms. Population abundance measurements of *Chloroflexus* from Ref. [45] show that similar distributions result for this species when constrained by a similar thermal gradient.

The results of modeling the total population density distribution, $\rho_t$, at equilibrium as a function of temperature are shown in Fig. 4, where the results of our model are compared with the experimental data. These results were computed using (7) with the parameter values explained above, the starting population densities defined by parabolic distributions determined by the birth rate dependencies of each strain, and the experimentally-determined predator distribution from Fig. 3. We passed the results of these calculations through a 50 point smoothing FFT filter (giving a 5 $^{\circ}$C averaging interval) to reflect the typical range of short-term thermal variation; this had the effect of smoothing abrupt changes at the edges of distinct population distributions. For these simulations, the only free parameter was the competition factor for predation, $\mu_p$, in (7), which was varied to obtain agreement with the measured population density as a function of temperature. The computed total population distributions agreed well with results from Refs. [22-24, 45] for values of $\mu = 0.004 \pm 0.0002$ (in agreement with the value obtained from the argument for the optimal temperature distribution) and $\mu_p = 0.025 \pm 0.005$.  These choices of parameter were highly constrained by the empirical data, since higher values of $\mu$ resulted in population competition so strong that regions within the observed microbial mats were reduced to zero, while the value of $\mu_p$ was determined to within this interval by the observed lower temperature bound of microbial mats.  For comparison with the DGGE data that provides maps of the distribution of distinct temperature strains across the gradient (Fig. 5(a)), we also computed the simulated distribution of different temperature strains along the thermal gradient for two different sets of temperature strains.  In Fig. 5(b), results of simulations are shown for the temperature strains plotted in Fig. 5(a), while Fig. 5(b) was generated using the birth rate data from the temperature strains in Fig. 2(b), which include a low temperature strain not observed in the DGGE data.

## 2. Effects of temperature variation

Seasonally, both sunlight intensity and ambient temperature vary appreciably at the geographic locations at which the field studies we wish to model were performed.  However, the observed adaptation to reductions in intensity observed in the studies considered here provides additional reason to believe we do not need to account for variations in sunlight from season to season.  Our review of the literature on the effect of illumination on photosynthesis revealed little dependence on natural ranges of annual variation in sunlight intensity. [6, 39, 44, 49]  Consequently, in our simulations we focused only on temperature variations throughout the year.  Data from seasonal observations of strain abundances and distributions show significant differences between fall, winter, spring and summer, although the inferred total population abundances for *Synechococcus* are similar for spring and summer.

Using this information, we examined the effect of both gradual (e.g., seasonal) variation in temperature as well as abrupt changes; these regimes are distinguished by comparing the timescales for temperature changes with the average bacterial birth rates. Fig. 7 shows the results of using (7) to model the total population densities as a function of time, summed over the entire thermal gradient. For an annual temperature variation with an amplitude of $10^{\circ}$C, corresponding to measured values for this latitude from Ref.[39], we computed the equilibrium values and total variation in population densities using (1), fixing $\mu = 0.004$, $\beta_o = 0.583$, and varying $N_s$, as described above, to determine whether different numbers of temperature strains offered advantages in a fluctuating thermal environment. In Fig. 7 (a) we show the dependence on diversity of the time-average total population for $10^{\circ}$C fluctuations normalized by the value in the absence of thermal variation: $\left\langle \rho_t \left(10^o C\right)\right\rangle \big/ \rho_t \left(0^o C\right)$, since for fixed $\mu$ there is some variation with $N_s$. There remains a 10% non-monotonic dependence on $N_s$ that does not favor increasing the number of strains; without normalization, this same trend holds though the variation is ~30%. Fig. 7(b) shows the result of computing the percent amplitude of population variations in total population, $\Delta\rho \big/ \left\langle \rho_t \left(10^o C\right)\right\rangle$, again for $10^{\circ}$C fluctuations; although there is a drop from $N_s = 2$ to higher values, there is little variation between results for $N_s = 3,4$ and 5. These simulations indicate that this model at least does not support the argument that higher diversity significantly enhances the stability of an assemblage of populations exposed to a fluctuating thermal environment throughout the year.

We also performed simulations to model the conditions found in experiments on the dynamics of bacterial growth by Meeks and Castenholz [43], which reported on cell culture studies of the growth response of *Synechococcus* cultures exposed to abrupt temperature shifts. Their samples were obtained from water > 70°C at Hunter's Hot Spring, Oregon, without further genetic sequencing or separation into distinct temperature strains; the cultures obtained were thus likely to consist of a mixture of high temperature strains. In Ref. [43], bacterial population growth (monitored by both cell counts and optical density) was measured for thermal shift experiments in which cell cultures were grown at a starting temperature for ~36 hours, then abruptly changed to a new temperature for several days. For temperature jumps from 55 °C to 65 °C, 65 °C to 55 °C and 65 °C to 45 °C , exponential growth held both before and after the temperature shifts, with a lag period following the step in temperature. (Fig. 8 (a)) We simulated a similar scenario by assuming a starting distribution of thermal strains with varying temperature optima, using birth rate vs. temperature data from the same temperature strains, as reported in Ref.[29] and no competition factor (to reflect the dilute solutions used). After being equilibrated at 55 °C for 36 hours ($\Delta t = 1$ hr), this system was then subjected to an abrupt temperature increase to 65 °C; the resulting evolution of population density as a function of time is plotted in Fig. 8 (b). Both laboratory and simulated datasets were fit to exponential growth models to extract average birth rates for each equilibration temperature.

# V. DISCUSSION

Our analysis of previous data from the literature and accompanying simulations have revealed that a relatively simple model can indeed describe both the distribution of optimal temperatures and the temperature dependence of cyanobacterial population densities. The finding that there is some optimal spacing $\Delta T$ and number of temperature strains $N_s$ for each value of the competition factor, $\mu$, is easy to explain using our model. In the absence of competition ($\mu = 0$), the number of temperature strains can proliferate without constraint ($\Delta T \rightarrow 0$) even while occupying overlapping temperature ranges, the only limit being the carrying capacities. In that case, the greater the number of temperature strains, $N_s$, and the smaller the temperature difference, $\Delta T$, the greater the total bacterial population, $\rho_t \propto N_s$. However, in the presence of competition ($\mu \neq 0$), the optimal population density for fixed $N_s$ is achieved by spacing out the different temperature strains uniformly, since temperature strains sharing even part of their range will compete and reduce their combined equilibrium population. This argument predicts that for each value of $\mu$ there will be some optimal value of $N_s$ (and a corresponding value of $\Delta T$) for which the total population is at a maximum. This is illustrated in Fig. 6, where results from our model for the total population are plotted for various fixed values of $\mu$, the only adjustable parameter in our model, and $N_s$. These total populations, $\rho_t$, were computed using (1) and birth rate and photosynthetic rate parameters drawn from fits to (3) and (4) using data from Fig. 2. For each value of $\mu$, there is indeed an optimum number of strains which agrees with the observed values of $N_s = 4$ and $\Delta T = 5.2 \pm 0.7\ ^{\circ}C$ for $\mu = 0.0046 \pm 0.008$. We consequently used that value of the competition factor in our remaining calculations for population density.

A comparison of the total population density of photosynthetic bacteria as a function of thermal gradients for four different locations revealed similar distributions independent of specific genetic makeup, spatial size of the gradient, specific chemical composition of the hot spring or geographic location. (Fig. 4) While the predominant factors in determining this distribution's shape are the limiting temperatures due to the lethal temperature and predators at lower temperatures, the correct functional dependence was only yielded in our simulations when empirical values of the birth rates, photosynthetic rates and specific choices of the competition factors were used.

Although this data averages over different temperature strains, in one case [42] the researchers were able to break this down into distributions of individual temperature strains. (Fig. 5) We were able to compare both sets of data to results from our simulations, accomplished by using measured values for the temperature dependence of birth rate and photosynthetic rate for carrying capacity. (Fig. 2(b)) Using the value of competition factor, $\mu$, determined from the calculation of optimal number of strains, we obtained a high degree of agreement between the simulated and measured total population temperature profiles, as well as with profiles for the individual strains, by varying the one remaining free parameter, the predator competition factor, $\mu_p$, which largely determines the low temperature behavior. Other *Synechococcus* strains with overlapping thermal regimes were not detected in the DGGE analysis and hence were not modeled in our simulations; the likely explanation is that these strains are adapted to different illumination conditions and hence are not seen in this season. [44] We also ran a separate simulation that replaced strain B' with B''', the lowest temperature strain from Ferris and Ward's study and one that was not detected in their DGGE data. [44]. At equilibrium, we found that this strain's population is preyed upon so efficiently by the ostracods and outcompeted by the other strains that its total population is only 7.7% of the next least populous strain. Thus, the fact that its presence is not detected in the DGGE data is not surprising.

To show that our model could also account for laboratory studies of growth behavior as a function of temperature, we used our model to simulate the same conditions as the cell culture experiments by Meeks and Castenholz. [43]. Our results exhibited the same behavior of initial and final exponential growth, with an intervening lag period where the populations of different strains adjusted to the new temperature. (Fig. 8 (b)) Thus, the basic functional behavior observed in Fig. 8(a) is reproduced by our model, and the ratios of final to initial birth rates agreed well ($2.2 \pm 0.2$ for the experimental results vs. $1.9 \pm 0.01$ for our simulations); as expected, the exact values of birth rates were not in agreement because of differences in illumination and specific strains used in their study and in the reference from which our parameters were drawn. In our simulations the intervening lag period was measured in only a few time steps ($\sim 3$ hours) as the populations readjusted, not $\geq 15$ hours as in the experiments, pointing to the action of some physiological adaption not present in our model as the dominant mechanism for this transient phenomenon.

The studies presented here consider only highly simplified models that attempt to capture the dominant influence of the physical environment. These microbial mats have significant structure, both vertically and with the plane of the mat. In many settings, different bacterial populations live in close juxtaposition, striated vertically, like *Chloroflexus* and *Synechococcus*, or intermingled. The available sunlight varies vertically, both in terms of intensity and the spectral content. It is possible that different strains of a single species might have different adaptations to illumination, as well as temperature.[28] Although the assumption of a 1D thermal gradient accurately captures the important dependence of community structure, actual effluent channels also exhibit V-shaped bands of different bacterial populations that reflect the temperature profile of the hot spring outflow as it is cooled by exposure to the air and ambient temperatures on each shore. Bacteria in these systems can develop symbiotic relationships in which neighboring organisms utilize the oxygen and chemical species excreted by the phototrophs, but such interactions are not considered here. [5][14] We also have used similar competition factors for both different species and different strains of the same organisms, in the absence of information about how to compute differences. Finally, this model uses actual field data for only one predator species to model predation, although the details of this interaction could vary substantially for other metazoans that graze upon the mats.

## VI. CONCLUSIONS

Our model shows that the current hot spring thermal gradient ecosystems can be explained successfully by population biology models that incorporate information about the physical environment and bacterial biophysical parameters; however, it does not account for the evolutionary mechanisms that established such a distribution of traits to begin with. As a consequence, a possible future research program could use artificial life models to include a simplified genome with "genes" that determine, e.g., $T_o$.  Such a model could be used to study  the way in which environmental factors influence the evolution of thermotolerance from an original temperate population, testing the interpretation by Miller and Castenholz [29] of their phylogenetic data.  One could include subpopulations in which the optimal temperatures for photosynthesis and birth rate assume a variety of values, and natural selection determines the eventual population distribution and its dynamics.  Earlier work by Bull and Fogarty illustrates how similar artificial life models have been used to explore topics in endosymbiosis and the evolution of multicellularity. [50, 51] We have conducted preliminary studies using such a model, and found that more widely distributed values of $T_o$ were favored relative to models with strains with too closely spaced a distribution of $T_o$; the latter case resulted in closely-spaced strains being outcompeted by more widely separated strains.

Additional experimental testing of our population biology-based models could occur via the creation of laboratory systems that recreate the environmental conditions considered.  It is interesting to imagine experimental designs in which these models, and their implications for evolution, are investigated through laboratory realizations of physical environments in the past and present such as salinity, light, oxygen and sulfide or thermal gradients [52-54].  Evidence from at least one study indicates that naturally-occurring distributions of thermophiles can be reproduced in a man-made thermal gradient [52].  The possibility of testing the generalization of this model to salinity gradients is especially interesting because the dependence of birth rate and photosynthetic rate on salinity follows a relationship similar to that with temperature [55-57], while abundances of cyanobacteria with different salinity preferences have in fact been mapped along a salinity gradient [32].

Nold and Ward [58] note that microbial mat communities do not appear to be actively growing under ordinary conditions, although they are undergoing photosynthesis.  Thus, our results might only model the conditions that correspond to the establishment of initial population distributions.  The case of dynamic changes in such populations has not been explored in conjunction with ways of separately culturing distinct temperature strains.  Even so, data from Meeks and Castenholz on the effect of stepwise changes in temperature agree well with the functional behavior of our model. [43] Experiments based on either natural or laboratory studies of sudden heating or cooling events would be illuminating.

Norris *et al.* [59] have also examined the case of a geothermal heating event and its consequences for species diversity.  They also performed laboratory experiments to simulate this event for comparison, observing that the populations underwent radical readjustments over approximately one week, then stabilized at new values in approximately 2 to 3 weeks.  Clearly this took place within the context of existing, randomly dispersed populations coming into a new equilibrium, rather than through mutations and the creation of new strains.  This could be modeled by introducing new species with different temperature preferences at random during the time series.

Future work might include incorporation of the effect of light intensities into our model.  For example, there is considerable data concerning the vertical structure of these mats, including the intensity and spectra composition of sunlight as a function of distance, and the population distribution and rate of photosynthesis by different organisms at different depths within the mat.  Extending these models to include a two dimensional slice through the mat that incorporates the effect of variation in both temperature and sunlight should be both interesting and feasible.

## ACKNOWLEDGEMENTS

We would like to acknowledge useful conversations with Robert Manning and Ted Wong.  This research was supported by a grant from the Howard Hughes Medical Institute Undergraduate Science Education Program and by the Andrew W. Mellon Foundation through a Tricollege New Directions Fellowship award to S.Amador Kane.

## CAPTIONS

FIG. 1  (Color online) Aerial view of Grand Prismatic Spring (Yellowstone National Park) illustrating bands of color due to different bacterial populations living in microbial mats along a thermal gradient.  Jim Peaco, July 2001,Yellowstone National Park Image by NPS Photo.

FIG. 2  (Color online) (a)  Thermal gradients in hot spring effluents often display populations of different strains of cyanobacteria with varying optimal temperatures for reproduction and photosynthesis.  Schematic diagram showing the characteristic dependence of either birth rate or photosynthetic rate on temperature, $T$, as well as the definition of $\Delta T$, the difference between optimal temperatures of thermal strains, and $T_H$ and $T_L$, the highest and lowest temperature for nonzero birthrates for each strain.  Model birth rate curves were generated using the parabolic models in (2) and (3).  (b)-(d)  Birth rate ((b) and (c)) and photosynthetic rate (d) vs. temperature data for different

temperature strains along thermal gradients in hot spring effluents, replotted from the

references indicated: (b) *Synechococcus* from Hunter's Hot Springs in Oregon, sampled

at 45, 48, 53, 55, 60, 66, 71 and 75 $^{o}$C, from Ref. [46];  (c)  *Synechococcus* strains  A, A', B,

B''' from Ref. [44] (Data from two strains, B' and B'', were omitted because they had the

same optimal temperature as strains B and A, respectively.);  (d)  *Chloroflexus* from

Lower Geyser Basin sampled at 45, 50, 60 and 72 $^{o}$C.  from Ref. [45].

FIG. 3

Measured population density (circles), $\rho_o$, of the ostracods, predatory crustaceans that graze on cyanobacteria microbial mats, as a function of temperature along a hot spring thermal gradient, replotted from Fig. 7 of Ref.  13.  A polynomial fit was used to model the measured distributions in our simulations (solid line);  a linear fit and quartic-polynomial fits were used, respectively, for temperatures below and above 47.5 $^{o}$C.

FIG. 4  (Color online)  Total population density, $\rho_t$, summed over all temperature strains

for the photosynthetic bacterium *Synechococcus* from various hot springs (lines and

symbols) and simulations (line only).  Data were replotted from the references indicated:

spectrophotometric measurements of BChl *a* absorbance (squares) Lower Geyser Basin

(Mushroom Springs), Yellowstone National Park. [22] and (circles), Lower Geyser Basin

(Mushroom Springs), Yellowstone National Park [45]; (triangles up) Thailand (replotted

from data from Table 3 in Ref. [24]);   (triangles down) DGGE intensity for all strains,

Octopus Springs, Yellowstone National Park from Fig. 1(a) in Ref. [23]; (solid

line):simulated equilibrium values of $\rho_t$ vs. $T$, generated using (7) and the birthrate and photosynthetic rate data from Ref. [44] (Fig. 2(b)).

FIG. 5  (Color online) (a)  DGGE intensity for different temperature strains of

*Synechococcus* from an effluent channel thermal gradient in Octopus Springs,

Yellowstone National Park as a function of temperature from Fig. 1(a) in Ref. [23], for

strains A (square), B (circle), B' (triangle up), A' (triangle down).  (b)  Simulated

distribution of the strains B, B', A and A' from (a) obtained using (7) and  the measured

birth rate and photosynthetic rate vs. $T$ data for these strains from Fig. 2 (b) and Ref. [44]

including ostracod predator interactions.

FIG. 6   Results of the simulation for the total population summed over all strains, $\rho_t$,

with each curve corresponding to varying numbers of strains, $N_s$, and fixed competition

factor, $\mu$.  For each curve, there is an optimal value of $N_s$ for which the total population is

at a maximum for that value of $\mu$:  (squares)  $\mu = 0.0055$, $N_s = 3$;  (circles) $\mu = 0.0046$, $N_s$

$= 4$; (triangles up) $\mu = 0.003$, $N_s = 5$;  (triangles down) $\mu = 0.0025$, $N_s = 6$;  (diamonds) $\mu$

$= 0.0020$, $N_s = 7$.

FIG. 7  Results of simulations to model the effects of seasonal temperature variation.

Equation (7), using parameters as described in the text, was used to compute the total population density, $\rho_t$ , summed over the entire thermal gradient, as a function of time when exposed to sinusoidal temperature variations of varying magnitude.  An amplitude of $10^{o}$C corresponds to measured values for the latitude at which most empirical studies were conducted.  (a) The total population density normalized to zero temperature fluctuations, $\left\langle \rho_t\left(10^o C\right)\right\rangle \big/ \rho_t\left(0^o C\right)$, was computed for varying numbers of temperature strains.  Increasing $N_s$ (more diversity) does not give a greater total population and hence greater fitness.  (b) The fractional peak-to-peak sinusoidal variation in total population, $\Delta\rho \big/ \left\langle \rho_t\left(10^o C\right)\right\rangle$ shows a decrease as the number of strains is increased beyond 2, but no stabilization against temperature variation as $N_s$ is increased beyond 3.

FIG. 8  Comparison of time evolution experiments for mixtures of different cyanobacterial temperature strains.  (a) Cell culture data (squares) replotted from Fig. 6 in Ref. [43] and (b) results from simulations (solid line) for laboratory experiments in which a mixture of *Synechococcus* high temperature strains were grown at 55 $^o$C, then increased to 65 $^o$C after 36 hrs.  Fits to exponential growth are shown as dashed lines and dotted lines (before and after the temperature increase).

## REFERENCS

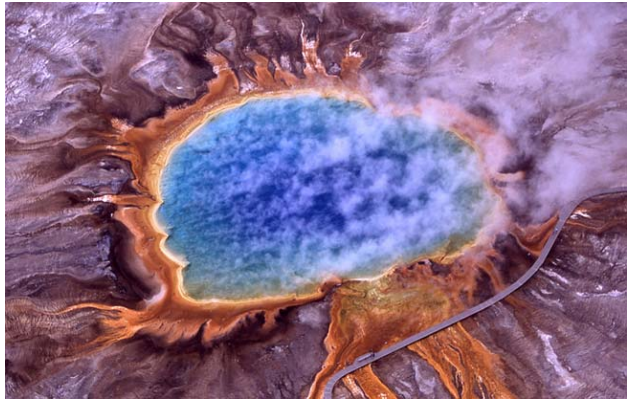

**Figure 1**

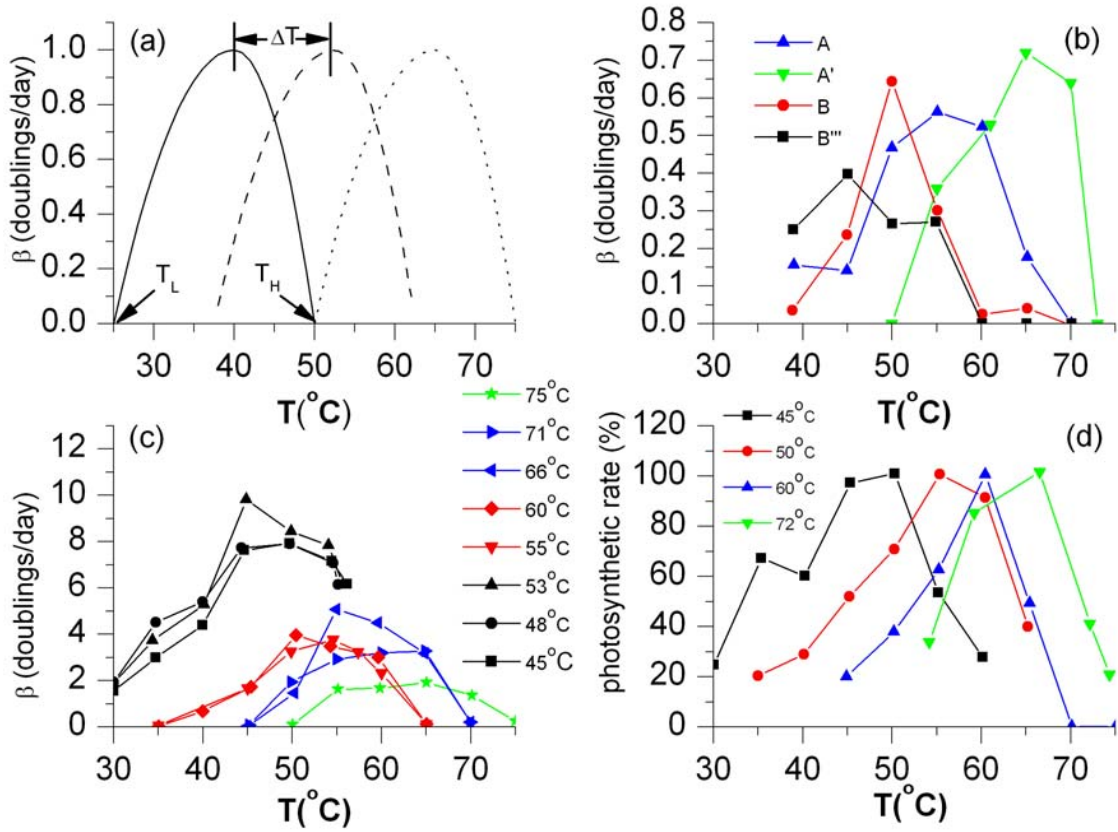

**Figure 2**

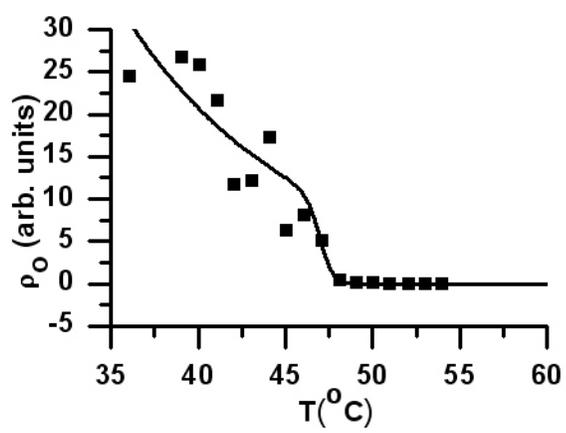

**Figure 3**

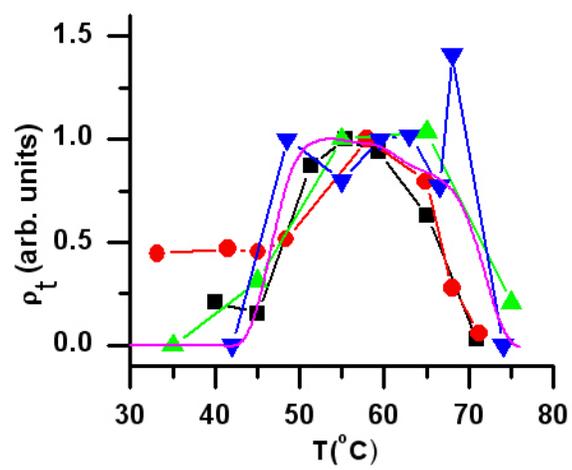

**Figure 4**

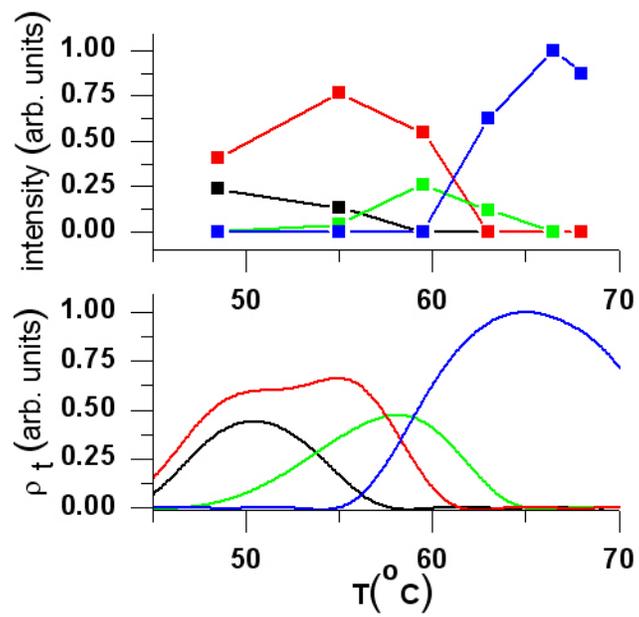

**Figure 5**

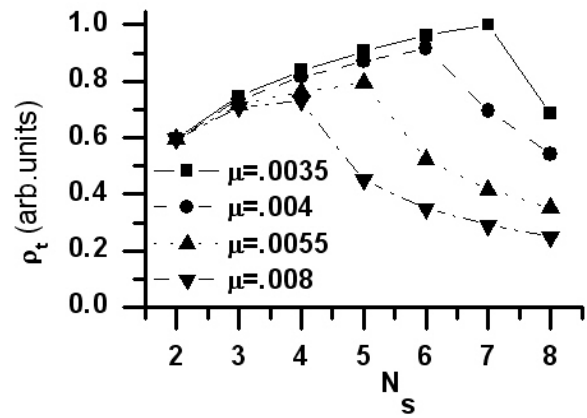

**Figure 6**

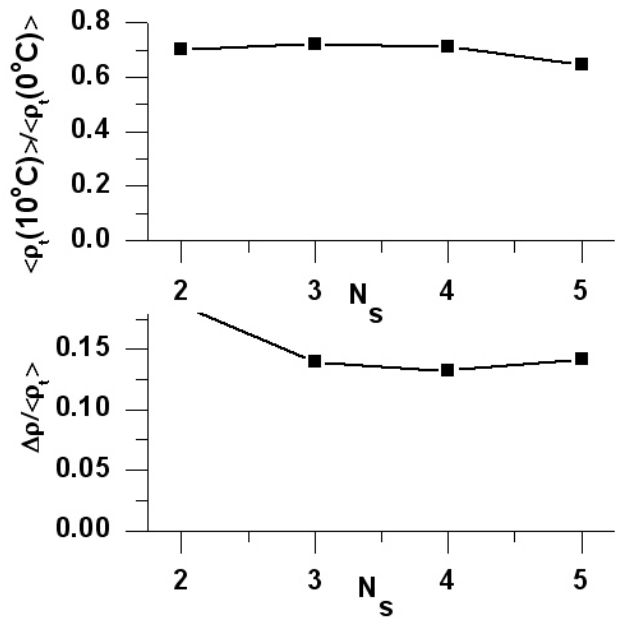

**Figure 7**

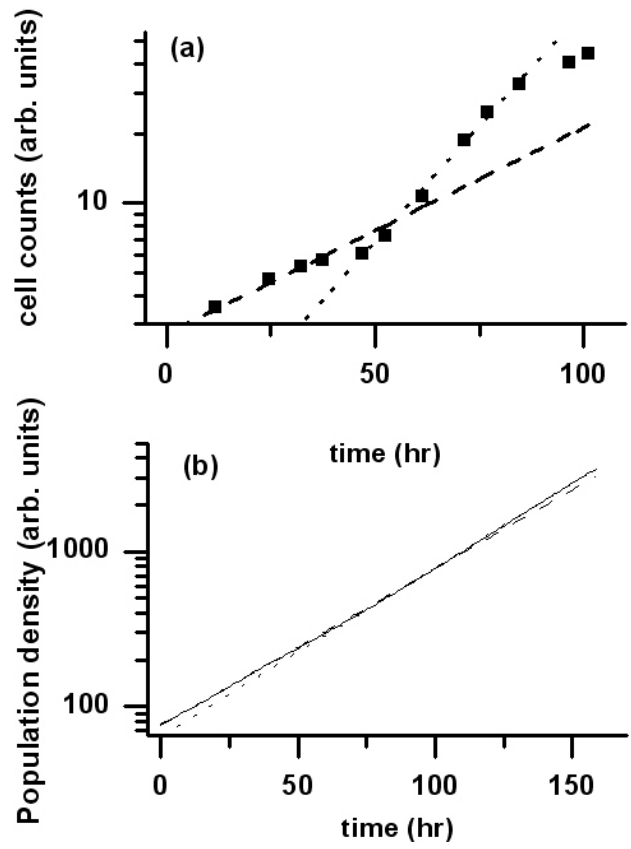

**Figure 8**